\documentclass{aa}  
\usepackage{amsmath}
\usepackage{amssymb}
\usepackage{graphicx}
\usepackage{color}
\usepackage{xcolor}
\usepackage{txfonts}

\newcommand{\msyr}{{\rm m\,s^{-1}\,yr^{-1}}}
\newcommand{\cmsyr}{{\rm cm\,s^{-1}\,yr^{-1}}}
\newcommand{\yr}{{\rm yr^{-1}}}
\newcommand{\kms}{{\rm km\,s^{-1}}}
\newcommand{\ms}{{\rm m\,s^{-1}}}
\newcommand{\cms}{{\rm cm\,s^{-1}}}
\newcommand{\nm}{{\rm nm}}
\newcommand{\lya}{Lyman-$\alpha$~}
\newcommand{\hi}{H~\textsc{i}~}
\newcommand{\lyb}{Lyman-$\beta$~}
\newcommand{\lcdm}{$\Lambda$CDM~}
\usepackage{siunitx}
\usepackage{booktabs}
\usepackage{longtable}

\begin{document} 

   \title{The ESPRESSO Redshift Drift Experiment\thanks{Based on Guaranteed Time Observations collected at the European Southern Observatory under ESO programmes 110.247Q and 112.25K7 by the ESPRESSO Consortium.}}

   \subtitle{III. The Third Epoch of QSO J052915.80-435152.0}

   \author{A. Trost \inst{1,2,3,4}\thanks{andrea.trost@inaf.it}, 
          C. M. J. Marques \inst{5,6, 7}, 
          S. Cristiani\inst{2,4,8}, 
          G. Cupani\inst{2,8}, 
          S. Di Stefano\inst{2,3},
          V. D'Odorico\inst{2,8}, 
          F. Guarneri\inst{2, 9}, 
          C. J. A. P. Martins\inst{5,6}, 
          D. Milakovi\'c\inst{2,8}, 
          L. Pasquini\inst{10,11},
          R. G\'enova Santos\inst{12,13}, 
          P. Molaro\inst{2,8}, 
          M. T. Murphy\inst{14},
          N. J. Nunes\inst{15,16}, 
          T. M. Schmidt\inst{17},
          Y. Alibert\inst{18}, 
          K. Boutsia\inst{19}, 
          G. Calderone\inst{2},
          J. I. Gonz\'alez Hern\'andez\inst{12, 13}, 
          A. Grazian\inst{20}, 
          G. Lo Curto\inst{10}, 
          E. Palle\inst{12,13}, 
          F. Pepe\inst{17}, 
          M. Porru\inst{2},
          N. C. Santos\inst{6,7}, 
          A. Sozzetti\inst{21}, 
          A. Su\'arez Mascare\~no\inst{12 ,13}, 
          M. R. Zapatero Osorio\inst{22} 
          }
    \institute{\textit{Affiliations are listed at the end of the paper.}}
   \date{Received XXX; accepted YYY}
  \abstract
{The Sandage–Loeb (SL) test probes cosmic expansion directly by measuring the redshift drift in quasar absorption features in a model-independent way. In this series of papers, we have launched an observational campaign to  assess whether current instrumentation is capable of measuring this effect and what systematic effects might interfere with a detection.}
{
We report the observations and analysis of the third epoch of ESPRESSO observations of the bright quasar J052915.80-435152.0 (SB2, $z_{\rm em} = 3.962$, $i=16.071$ mag), extending the temporal baseline to $\sim2$ years, assessing systematics and measurement procedures, and providing the tightest constraints on the redshift drift in the series so far.
}
{
We acquired 9.5 hours of ESPRESSO observations, complementing the 12 hours presented in the first paper of the series, with one year of separation from the second epoch. The complete dataset, composed of three separate and independent epochs, was analysed and compared to spline-based \lya forest models calibrated on simulations, to measure the presence of any velocity drift among the spectra. The measurement was carried out with two independent methods, Pixel-by-pixel and Likelihood correlation.
}
{
Both approaches give a consistent null result, 
$\dot{v} \approx -3.5 \pm 3.6 ~\msyr $
(or $\dot{z} \approx (-5.3\pm5.6)\times 10^{-8}~{\rm yr^{-1}}$ in redshift space), in agreement with \lcdm expectations ($\dot{v}_{\Lambda \rm CDM}(z=3.57)=-0.41~\cmsyr$), systematic effects remain subdominant at the present level of noise. By extrapolating the results from the observed sightline to the complete QUBRICS Golden Sample, we show that ESPRESSO alone could detect the signal on century timescales, while a joint ESPRESSO+ANDES programme would reach first detection before 2080. A future analysis of the other quasars of the QUBRICS Golden Sample is required to improve this estimate. We show that the program would greatly benefit from a complementary effort with radio facilities targeting low-z \hi 21 cm absorption lines. Such synergy could reduce the experiments' timeline by up to $\sim 10$ years.
}
{}
\titlerunning{The ESPRESSO Redshift Drift Experiment III}
\authorrunning{A. Trost et al.}
   \maketitle
    \nolinenumbers
\section{Introduction}
The Sandage-Loeb test \citep[SL:][]{Sandage62, Loeb98} is a uniquely direct and model-independent probe of cosmic dynamics, offering a real-time measurement of the Universe’s expansion rate by tracking the slow temporal drift in the redshift of distant sources. In an era where traditional geometric probes, such as Type Ia supernovae, Baryon Acoustic Oscillations (BAO), and the Cosmic Microwave Background (CMB) have achieved remarkable precision \citep{Abbott2024, DESI2024, Tristam2024, Madhavacheril2024}, the SL test provides a powerful and complementary approach. By providing a direct, empirical measurement of the redshift's rate of change, $\dot{z}$, the SL test bypasses reliance on standard candles or rulers. It offers a fundamental, model-agnostic test of cosmic dynamics, providing a direct constraint that any viable cosmological theory, from standard Friedmann–Lemaître–Robertson–Walker (FLRW) models to modified gravity, must satisfy \citep{Uzan, Liske08, ANDES}. 
In a standard FLRW Universe, the redshift drift effect takes the simple analytical form 
\begin{equation}\label{eq:drift}
    \dot{z} = (1+z)H_0 -H(z),
\end{equation}
where $\dot{z}$ can be directly measured with spectroscopic observations, simply detecting the movement of spectral lines over time, independently of any cosmological assumptions underlying the analytical form of Eq.~\ref{eq:drift}.

Over the past decades, the SL test has evolved from a theoretical curiosity into a viable observational frontier \citep{Corasaniti2007, Geng2014, Alves:2019hrg, Marques1}.
This transition from theory to practice is driven by immense technological progress. The primary observational challenge lies in measuring the exceedingly small signal, on the order of $\dot{z}\sim10^{-11}~\yr$, or $\dot{v}=c\,\dot{z}\,(1+z)^{-1}\sim0.5~\cmsyr$ in spectroscopic terms. However, the advent of ultra-stable, high-resolution spectrographs like ESPRESSO on the VLT \citep{Pepe20} and the upcoming ANDES instrument on the ELT \citep{Marconi2024, ANDES} is finally bringing a statistically meaningful detection within reach. Parallel efforts are also exploring the potential of SKA-like facilities to conduct redshift drift measurements using 21 cm absorption~\citep{Darling2012} or emission lines from neutral hydrogen at low redshift ($z\lesssim1$; \citealt{Rocha:2022gog, Marques2, Kang2025}).
  
In \citet[hereafter Paper I]{Trost2025b}, we reported the start of the first dedicated programme (i.e. not relying on archival data) aiming at carrying out the SL test using the Lyman-$\alpha$ forest of the most luminous quasar in the Universe, J052915.80–435152.0 (hereafter called SB2, $z_{\rm em}=3.962$, $i=16.071~\rm mag$), based on two high-resolution high signal-to-noise ESPRESSO spectra. The result, consistent with zero drift within uncertainties ($\dot{v}=-1.43 _{- 5.10} ^{+ 5.08} \; {\rm m\,s^{-1}\,yr^{-1}}$)
, met theoretical expectations and provided a realistic assessment of current observational capabilities. 
A complementary study (Marques et al, subm.; Paper II) has been carried out on another bright high-redshift quasar, J212540.96-171951.32 (SB1, \(z_{\rm em}=3.897\), $i = 16.548~\rm mag$), finding consistent results.

This work follows the previous papers, extending the experiment with a third observational epoch of quasar SB2, lengthening the temporal baseline, increasing data signal-to-noise (S/N) and improving measurement sensitivity. Specifically, we aim to:
\begin{enumerate}
    \item Validate and refine the modelling and measurement procedures developed in Papers I and II;
    \item Tighten constraints on the redshift drift signal through improved data and extend the experiment's temporal baseline;
    \item Diagnose and mitigate potential systematic effects that might become dominant at higher S/N;
    \item Update forecasts for present and future facilities.
\end{enumerate}
This paper is structured as follows: in Sect.~\ref{sect:data_acq} we summarise the observations and the data treatment. Section~\ref{sect:liske_exp} estimates the expected result at the current S/N from literature scaling relations, whereas in Sect.~\ref{sect:drift_meas} we develop a model of the \lya forest of our target, and employ two methods to carry out the actual measurement of the redshift drift on the acquired data. In Sect.~\ref{sect:systs} we investigate the presence of systematic effects of either astrophysical or instrumental nature in our results, whereas Sect.~\ref{sect:forecast} reports an extrapolation of our results in the context of a greater observational effort, and estimates the total required experiment timeline. Section~\ref{sect:results} summarises and discusses our findings.
Thoughout the paper we assume an underlying $\Lambda\rm CDM$ cosmology with the following parameters: $\Omega_{\rm m}=0.308$, $\Omega_\Lambda=0.692$, $h=0.678$, $\Omega_{\rm b}=0.0482$, $\sigma_8=0.829$, $n=0.961$, and a helium mass fraction $Y_{\rm p}=0.24$. 

\section{Data acquisition and treatment}\label{sect:data_acq}
High-resolution, high-fidelity, stable spectroscopic observations of SB2 were carried out with ESPRESSO \citep{Pepe20} at VLT in single UT mode, in December 2024 (ESO period P114). We adopted the same instrumental setup, observing strategy, exposure time and calibration plan as those of observations carried out in periods P110 and P112, the first two epochs, and presented in Paper I. 
The application of exactly the same instrumental setup is essential to ensure direct comparability between spectra, with the goal of measuring the redshift drift and reducing systematic effects. 
To summarise, the data have a resolving power of $R\sim135000$. A detector binning 4x2 was adopted to enhance S/N in the \lya forest, where ESPRESSO's efficiency is lower.
The observations are summarised in Table~\ref{tab:observations}, together with the two datasets relative to the first two epochs presented and analysed in Paper I.

The data were reduced with ESPRESSO Data Reduction Software v3.3 (DRS, \citealt{DiMarcantonio2018}). Both Fabry-P\'erot etalon (FP) + ThAr lamp and Laser Frequency Comb (LFC) frames were used for wavelength calibration of the spectra, producing two distinct datasets that were used to search for systematic effects in the wavelength calibration procedure (see Sect.~\ref{sect:wavelength}). Since ESPRESSO's LFC has experienced technical difficulties during the runtime of the presented experiment, the main results reported in the paper are based on the FP+ThAr calibration, whereas the LFC-calibrated spectra are used as a comparison.

The final products of the pipeline were flat-fielded, blaze corrected, sky subtracted, and wavelength calibrated. The spectra were also corrected to the Solar System barycentre using standard pipeline procedures, employing INPOP13c ephemeris \citep{INPOP13}, taking into account also Earth’s precession.

Order-by-order spectra were optimally extracted from the detector frames. 
In contrast to the procedure adopted in Paper I, we performed the relative flux calibration by removing the instrument response function directly on the order-by-order spectra. This was done to ease continuum estimation and reduce systematics related to flux equalisation before the combination of the spectra. To compensate for spurious distortions in the spectra due to differences in instrumental response, atmospheric conditions and fibre positioning on the sky, all orders of each spectrum were equalised to match the median flux values of the orders of the first observation. Subsequently, a sky mask was applied to the single exposures while retaining an order-by-order format, shifted by the specific barycentric correction (BERV).

\begin{figure*}
    \centering
    \includegraphics[width=\linewidth]{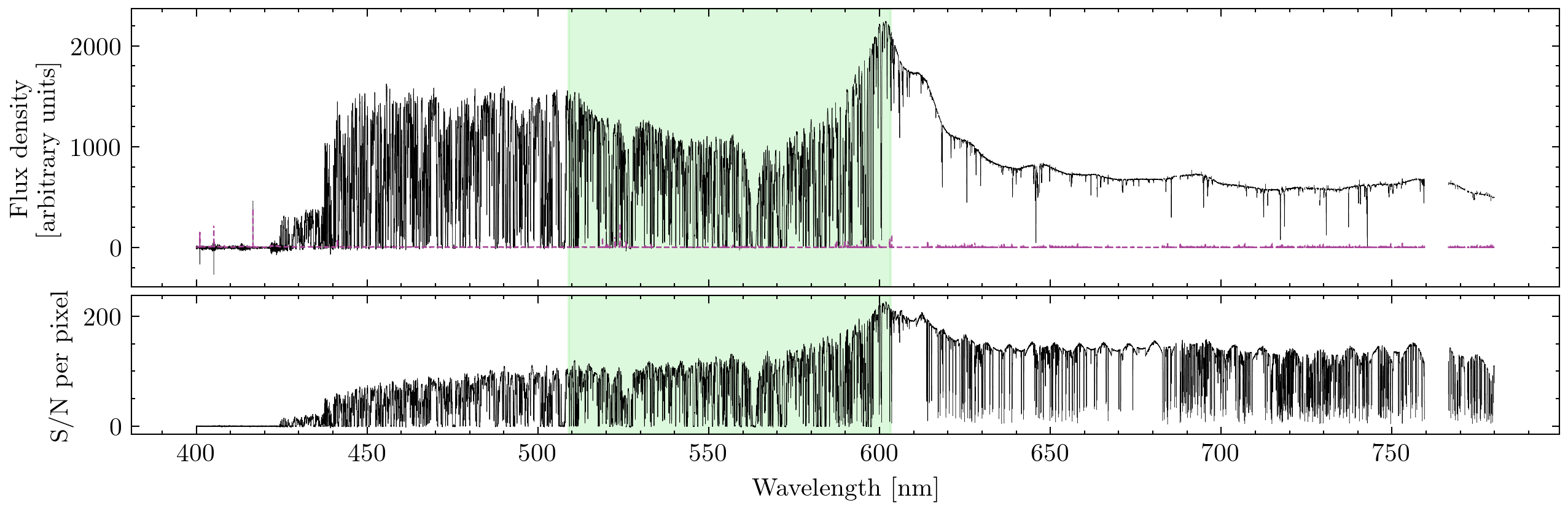}
    \caption{Combined spectrum of SB2. Top panel: Flux density in arbitrary units is shown in black, with the purple line denoting the flux density error. Bottom panel: S/N per $1~\kms$ pixel. The green shaded area highlights the \lya forest considered in the redshift drift measurement, bound by the \lya and \lyb emissions of the quasar, namely between 509 - 603 nm.}
    \label{fig:spectrum}
\end{figure*}

Lastly, the orders were merged and the exposures combined in three independent datasets using the \textsc{Astrocook}\footnote{\url{https://github.com/DAS-OATs/astrocook}} software \citep{Cupani20}, each comprising the exposures taken in each ESO period, producing the three epochs spectra. Another combined spectrum was produced by combining all exposures, and is shown in Fig.~\ref{fig:spectrum}. 
All combinations were produced by rebinning the different spectra all at once into a common final pixel grid with $1~\kms$ step, corresponding to the original detector's pixel size (when binned in the 2x mode along the dispersion axis) thus minimizing pixel correlations. 

The following analysis is based on the \lya forest region in these four spectra, spanning the range $509 - 603~\nm$, with median S/N per $1~\kms$ pixel at continuum of 47 (first epoch), 71 (second epoch), 75 (third epoch) and 113 (combined). For each epoch, an average date was computed by weighting the observation dates of the individual exposures by their S/N per $1~\kms$ at continuum (as defined in Table \ref{tab:observations}).
The time differences between the second and third epochs to the first are $\Delta t_2=0.87 ~{\rm yr}$ and $\Delta t_3= 1.96 ~{\rm yr}$, respectively.

Single-exposure order-merged spectra were also generated by rebinning the flux-calibrated, order-equalised, sky-masked order-by-order spectra to the common wavelength grid. These spectra have a median S/N per pixel at continuum in the \lya forest spanning the range 13-26, as reported in Table~\ref{tab:observations}.
In Sect.~\ref{sect:Obs} we investigate the effect of carrying out a measurement on the single exposures instead of on the combined epoch spectra.

\begin{table*}[t]
    \centering
        \caption{Summary of single observing blocks (OBs) of SB2 taken during observing periods P110, P112 and P114.  
}
    \begin{tabular}{lccccccccc}
    \hline
        Epoch & Period & ID & Exposure Start (UTC) & MJD & $t_{exp}$ & Grade & Seeing   & Airmass & S/N  \\
              &        &    &                     &     & [s]       &       & [arcsec] &         &      \\
    \hline
     1st & P110 & 1A & 2022-10-29 07:21:01 & 59881.31068001 & 3438.0 & C & 0.93 & 1.066 &18.6 \\
         &      & 1B & 2023-01-22 01:04:40 & 59966.04843679 & 3438.0 & A & 0.62 & 1.071&19.9 \\
         &      & 1C & 2023-01-22 02:13:07 & 59966.09453567 & 3438.0 & B & 0.72 & 1.075&17.4 \\
         &      & 1D & 2023-01-23 01:14:56 & 59997.04166665 & 3438.0 & C & 1.28 & 1.066&13.5 \\
         &      & 1E & 2023-02-22 00:54.53 & 59967.05436154 & 3438.0 & A & 0.60 & 1.116&22.8 \\ \hline
          
     2nd & P112 & 2A & 2023-08-25 08:41:12 & 60181.36712803 & 3438.0 & A & 0.42 & 1.278&22.0 \\
         &      & 2B & 2023-11-11 05:22:38 & 60259.22743581 & 3453.0 & C & 0.99 & 1.087&17.8 \\
         &      & 2C & 2023-11-19 05:02:35 & 60267.21411237 & 3453.0 & A & 0.64 & 1.078&22.7 \\
         &      & 2D & 2023-12-06 04:15:02 & 60284.18013713 & 3453.0 & A & 0.49 & 1.069&24.6 \\
         &      & 2E & 2023-12-07 04:41:30 & 60285.19820316 & 3453.0 & A & 0.56 & 1.065&23.6 \\   
         &      & 2F & 2023-12-11 04:47:35 & 60289.20252708 & 3453.0 & A & 0.57 & 1.070&25.3 \\
         &      & 2G & 2023-12-13 05:22:56 & 60291.22794520 & 3453.0 & A & 0.54 & 1.103&23.1 \\
         &      & 2H & 2024-02-05 04:15:20 & 60345.18147687 & 3453.0 & C & 0.50 & 1.512&17.6 \\ \hline
          
    3rd & P114  & 3A & 2024-12-26 01:05:00 & 60670.04514905 & 3453.0 & A & 0.54 & 1.215&21.4 \\
        &       & 3B & 2024-12-26 02:11:26 & 60670.09127605 & 3453.0 & A & 0.53 & 1.101&22.6 \\ 
        &       & 3C & 2024-12-26 03:18:08 & 60670.13759482 & 3453.0 & A & 0.54 & 1.065&21.8 \\
        &       & 3D & 2024-12-26 04:22:38 & 60670.18239184 & 3453.0 & A & 0.38 & 1.091&23.1 \\ 
        &       & 3E & 2024-12-27 01:10:19 & 60671.04883566 & 3453.0 & A & 0.73 & 1.194&18.8 \\
        &       & 3F & 2024-12-27 03:08:49 & 60671.13113179 & 3453.0 & B & 0.86 & 1.065&17.3 \\
        &       & 3G & 2024-12-29 02:59:01 & 60673.12431786 & 3453.0 & B & 0.70 & 1.065&16.0 \\ 
        &       & 3H & 2024-12-29 05:03:38 & 60673.21086665 & 3453.0 & A & 0.44 & 1.163&22.4 \\
        &       & 3I & 2024-12-30 01:29:23 & 60674.06208193 & 3453.0 & A & 0.48 & 1.137&22.9 \\ 
        &       & 3J & 2024-12-30 02:31:39 & 60674.10531465 & 3453.0 & A & 0.53 & 1.073&26.7 \\ \hline

    \hline

    \end{tabular}
    \label{tab:observations}
    \tablefoot{The columns of the Table indicate: i) epoch of observations; ii) ESO period; iii) OB identifier; iv) date of start of observations (UT time); v) date of start of observations (MJD time); vi) exposure time in seconds; vii) ESO OB grade, based on observational constraint compliance; viii) average atmospheric seeing during the exposure; ix) average airmass during the exposure; x) median S/N per $1~\kms$ pixel at continuum in the \lya forest.}
\end{table*}

\section{Expected uncertainty}\label{sect:liske_exp}
\cite{Liske08} defined a general scaling relation that estimates the average expected uncertainty on a velocity shift measurement carried out among an ensemble of many quasars. We adopt this relation as a first-order estimate of the expected uncertainty for the present dataset, extrapolating it to the case of a single target. We note, however, that sightline-to-sightline variations at the $\sim10-20\%$ level are expected due to differences in the number and clustering of \lya absorption lines (see Fig. 8 of \cite{Liske08})

The average velocity shift measurement uncertainty is expected to scale, in the context of an ESPRESSO Redshift Drift measurement, as
\begin{equation}\label{eq:liske}
    \sigma_v = 2\, g\,  \left(\frac{\rm S/N}{4058}\right)^{-1} \left(\frac{1+z_{\rm QSO}}{5}\right)^{-1.7} \left(\frac{N_{\rm QSO}}{30}\right)^{-1/2}f_{Ly\alpha}^{-1/2}~\cms,
\end{equation}
where $\rm S/N$ is the total $\rm S/N$ per 1$\kms$ pixel at continuum obtained over all exposures, $z_{\rm QSO}$ is the emission redshift of the quasar, while $N_{\rm QSO}$ is the total number of quasars considered in the measurement. The factor $f_{\rm Ly_{\alpha}}$ is the fraction of the Lyman-$\alpha$ forest pixels considered after masking metal lines, intervening galactic structures (e.g., DLAs) and bad pixels. In our case, these parameters take the values ${\rm S/N} = 113$, $z_{\rm QSO}=3.962$, $N=1$, and $f_{\rm Ly_{\alpha}}=0.77$.
 
The "form factor", $g$, found in Eq.~\ref{eq:liske}, depends on the observational strategy and the distribution of the single observations throughout the program. Assuming $N_e$ observational epochs, the $j$-th epoch being separated from the beginning of the program by $\Delta t_j$, we can define the fractional temporal baseline $h_j = \Delta t_j/ \Delta t$, where $\Delta t$ is the total temporal baseline of the experiment, 
and the fractional integration time $f_j$, i.e. the integration time fraction spent on the $j$-th epoch with respect to the total. Following Eq. 23 of \citet{Liske08}, the "form factor" can be defined analytically as:
\begin{equation}\label{eq:form}
    g(N_e, \textbf{h}, \textbf{f}) = \frac{1}{2}\left[ \sum_{j=1}^{N_e}h_j^2f_j - \left(\sum_{j=1}^{N_e}h_jf_j\right)^2\right]^{-1/2}.
\end{equation}
For the sequence of observations reported in Table~\ref{tab:observations}, $g=1.27$.

Based on this scaling relation, the expected precision in a redshift drift measurement with the current observations is $\sigma_v=5.7 ~\ms$ in velocity space. Accounting for the temporal baseline of the current experiment, the measurement uncertainty for the cosmological acceleration is $\sigma_{\dot{v}} = \sigma_v/\Delta t = 2.9 ~\msyr$, equivalent to $\sigma_{\dot{z}} = 4.4\times10^{-8} ~{\rm yr^{-1}}$. 

\section{Redshift drift measurement}\label{sect:drift_meas}
\subsection{Modelling the forest}\label{sect:model}
To measure a shift between spectral lines, we model the \lya forest in the spectrum of SB2 following the procedure described in Paper I. In brief, the combined final spectrum is fitted with 500 independent spline functions, $S_j(\textbf{n}_j)$, each defined by a unique node distribution, $\textbf{n}_j$. The knot array is generated starting from a equispaced distribution, $\textbf{n}_{j,0}$ within the boundaries of the \lya forest ($509 ~\nm \le\lambda\le 603~\nm$), with internodal spacing, $A_j$, randomly drawn from a split-normal distribution $\rm{SplitNormal}(A, \sigma^-_{A}, \sigma^{+}_{A})$, where $A$ and $\sigma^\pm_{A}$ are estimated from mock sightlines as described below. The positions of the spline nodes are then perturbed independently for each realization as $\textbf{n}_j = \textbf{n}_{j,0} + \textbf{m}_j$, where $\textbf{m}_j$ is a noise term randomly drawn from $\rm{SplitNormal}(0, \sigma^-_{A}, \sigma^{+}_{A})$.  

As for Paper I, we produced 100 mock spectra of the \lya forest of SB2 to calibrate the modelling procedure, extracting \hi optical depth along artificial sightlines piercing through the boxes of the Sherwood suite of hydrodynamical cosmological simulations \citep{Bolton17} at $3<z<4$, and stitching them to match the path length of SB2's \lya.
These sightlines were degraded to match the resolution, pixel size and S/N of the three observational epochs of our data \citep[following e.g. ][]{Rorai2017, Trost2025a}. The pixel mask accounting for metal lines, sub-DLA, telluric lines, and bad pixels considered on the SB2 data was also applied to the mocks. 

To mimic the cosmological drift, we introduced a small physical velocity drift to the gas particles in the simulations before computing the \hi optical depths, of order $-0.37 ~\cms$ and $-0.84~\cms$, in the second and third epoch's mock datasets, respectively. Each sightline has thus four realisations, simulating the first, second and third epochs, plus a combination of the three. 
The combined mock spectra were used to calibrate the best internodal distance $A=7.34~\kms$ and its uncertainties $\sigma^-_A={0.13}~\kms,\sigma^+_A=0.33$ that yield normally distributed residuals between the mock spectra and the fitted splines. Note that the estimated value of $A$ depends primarily on the $\rm S/N$ of the data, which is why it has been re-computed with respect to Paper I. It shows little to no sensitivity to the assumed cosmology or to the velocity shift applied to the mocks, provided the latter remains well below the pixel scale.

The final \lya forest model of SB2, $\bar{S}$, is taken as the average of 500 independent spline realisations, $S_j$. Fit residuals of the final model to the data are consistent with a Gaussian distribution with mean $\mu = 0.006$ and standard deviation $\sigma=0.979$, underling the adequacy of the method.
In the following analysis, outlier pixels with more than $5\sigma$ deviation from the average model were masked in all spectra. 

\subsection{Pixel-by-pixel method}\label{sect:pix2pix}
To measure the redshift drift that occurred between the three observational epochs, we build on the approaches presented in Paper I, the simplest of which is a direct application of the spectrum-to-spectrum comparison used in the context of radial velocity (RV) measurements for exoplanet searches, as defined by \cite{Bouchy2001}. This method was adapted to the redshift drift measurement and used in \cite{Liske08} to derive the scaling relation reported in Eq.~\ref{eq:liske}. 
However, unlike \cite{Liske08} and Paper I, in this work we exploit the model of the \lya forest transmission to evaluate the flux derivatives and compare the three epochs' spectra to the same common model, similarly to what is done with stellar RV using stellar spectral templates.

Summarising, the normalised flux at the $i$-th pixel of the $k$-th epoch spectrum, $F_{k,i}$, can be described as a first order perturbation of the \lya forest transmission model, $\bar{S}$, induced by a small velocity shift, $\delta v_{k,i}$, as:
\begin{equation}
F_{k,i} = \bar{S}(\lambda_i) - \frac{\partial\bar{S}}{\partial \lambda_i}\frac{\delta v_{k,i}}{c}\lambda_i,
\end{equation}
where each pixel of the spectrum defines a contribution to the velocity shift of the $k$-epoch
\begin{equation}
\delta v_{k,i} = \frac{\bar{S}(\lambda_i) - F_{k,i}}{\partial\bar{S}_i/\partial \lambda_i}\frac{c}{\lambda_i}.
\end{equation}
The velocity shift occurring between the model and the spectrum can then be estimated as a weighted average of the contributions of all pixels
\begin{equation}
\delta v_k  = \frac{\sum_i \delta v_{k,i}w_i}{\sum_i w_i}
\end{equation}
where the weights are defined as the inverse of the variance of the single pixel velocity contribution
\begin{equation}\label{eq:p2pweights}
w_i  = \sigma_{v_{k,i}}^{-2} = \left[ \frac{c}{\lambda_i \left(\partial\bar{S}_i/\partial \lambda_i\right)}\sigma_{F_{k,i}}\right]^{-2},
\end{equation}
where $\sigma_{F_{k,i}}$ is the error on the normalised flux at pixel $i$ and epoch $k$. 
Note that, differently from a spectrum-to-spectrum comparison, we compared the data to a transmission model and therefore Eq.~\ref{eq:p2pweights} neglects model uncertainties with respect to the observational flux errors. 

The uncertainty on the final $\delta v_k$ estimate is
\begin{equation}\label{eq:sdv_p2p}
\sigma_{\delta v_{k}} = \left[ \sum_i \sigma_{v_{k,i}}^{-2} \right]^{-1/2}.
\end{equation}

Applying this approach to the three observational epochs yields 
$\delta v_1 = -0.31\pm 6.44~\ms$, $\delta v_2 = 1.47 \pm 4.11~\ms$, and $\delta v_3 = -5.09\pm 3.91~\ms$. 

From the three estimates, we carried out the measurement of the cosmic acceleration by means of a linear fit in the $\delta v - t$ plane, with equation 
\begin{equation}\label{eq:dvdt_fit}
    \delta v = \dot{v}\cdot \, (t-t_0) + q,
\end{equation}
where the pivot point $t_0$ is the average observation date, weighted by the single epoch measurement uncertainties, and is adopted to minimize the covariance between the fit parameters. The estimated cosmic acceleration is the best fit slope
$\dot{v} = -3.43 \pm 3.56~\msyr$,
equivalent to a redshift drift of 
$\dot{z}= \left(-5.23 \pm 5.43 \right)\times 10^{-8} ~{\rm yr^{-1}}$.

\subsection{Model likelihood correlation}\label{sect:Likelihood}
An alternative approach to the measurement is carried out by maximising the likelihood between model and data, as a function of a velocity shift, $\delta v$, between the two. 
We stress that, following Eq.~\ref{eq:drift}, the redshift drift effect is redshift dependent, i.e. $|\dot{z}|$ is larger on the blue end ($\dot{v}(z=3.96)=-0.53~\cmsyr$) than on the red end ($\dot{v}(z=3.19)=-0.33~\cmsyr$) of the forest, thus introducing a compression into the spectral shapes of the \lya lines. Such an effect is negligible over the span of SB2's \lya forest at the current level of S/N of our data. The following analysis can be easily generalised to account for this compression effect in future high-S/N ($\gtrsim 1000$) measurements.
On these grounds, we considered as a good approximation a rigid shift of the model in velocity space in order to measure the drift occurring between epochs.

For each epoch $k$, we determined the relative velocity shift $\delta v_k$ between the model and the observed spectrum by maximising the likelihood function
\begin{equation}\label{eq:likelihood}
    \ln\mathcal{L}(\delta v_k) = -\frac{1}{2}\sum_{i=1}^{N_{\mathrm{pix}}} \frac{\left[F_{k,i} - \bar{S}(\lambda_i, \delta v_k)\right]^2}{\sigma^2_{F_{k,i}}},
\end{equation}
where $F_{k,i}$ and $\sigma_{F_{k,i}}$ denote the normalised flux and its associated uncertainty at pixel $i$ of the $k$-th epoch spectrum. The model $\bar{S}(\lambda_i, \delta v_k)$ represents the mean \lya{} forest transmission as constructed in Sect.~\ref{sect:model}, rigidly shifted in velocity space by $\delta v_k$ and evaluated on the same wavelength grid $\{\lambda_i\}$ as the data. The summation in Eq.~\eqref{eq:likelihood} is performed over all non-masked pixels within the \lya{} forest region.

Note that, differently from the method used in Paper I (Sect. 2.4, Eqs. 11-12), we abandoned the idea of weighting pixels based on the model derivative, as introducing the weights underestimates the measurement uncertainties, and relied on a more robust statistical estimator. 

We evaluated Eq.~\eqref{eq:likelihood} over a dense grid of velocity shifts in the range $-50~\ms \leq \delta v_k \leq 50~\ms$ and fitted a parabola to the resulting $\ln\mathcal{L}(\delta v_k)$ profile to obtain a precise estimate of its maximum. The statistical uncertainty on $\delta v_k$ was derived from the curvature of the log-likelihood at the maximum,
\begin{equation}\label{eq:sdv_like}
    \sigma_{\delta v_k} \approx \left[-\left.\frac{d^2 \ln\mathcal{L}}{d (\delta v_k)^2}\right|_{\delta v_k,{\rm max}}\right]^{-1/2},
\end{equation}
following standard maximum likelihood theory.

The comparison between the model and the three spectra yields $\delta v_1 = 0.51\pm6.59~\ms$, $\delta v_2 = 2.78\pm4.21~\ms$ and $\delta v_3 = -4.39\pm4.00~\ms$. 

Again, these three estimates were used to measure the cosmological signal, by means of a linear fit in the $\delta v -  t$ plane of Eq.~\ref{eq:dvdt_fit}, yielding a cosmic acceleration of $\dot{v} = -3.63\pm3.65~\msyr$, equivalent to a redshift drift of $\dot{z} = \left(-5.53 \pm 5.56\right)\times 10^{-8}~\yr$.

\begin{figure}[t]
    \centering
    \includegraphics[width=\linewidth]{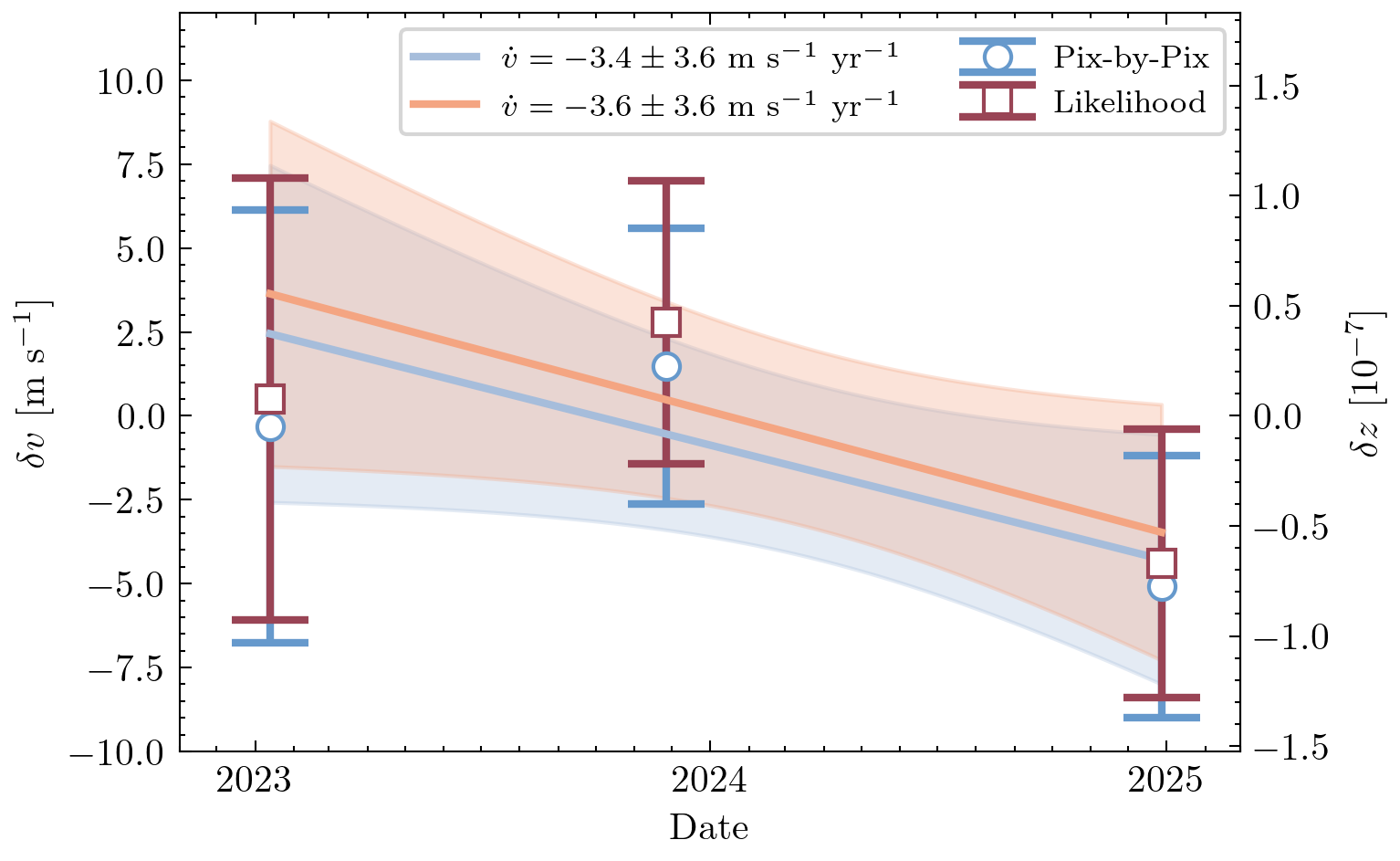}
    \caption{Velocity shifts of every single epoch with respect to the combined mean model, as a function of time, obtained via the Pixel-by-pixel (Sect.~\ref{sect:pix2pix}, blue circles) and the Maximum Likelihood approaches (Sect.~\ref{sect:Likelihood}, red squares). For both approaches, the cosmic acceleration is computed as a linear fit of the data points and shown as a solid line. Shaded areas report the uncertainty in the fit.}
    \label{fig:Drift_dvdt}
\end{figure}

Fig.~\ref{fig:Drift_dvdt} shows the velocity shifts between the transmission model and each spectrum obtained via the two independent methods, as well as the best fit linear relation passing through the datapoints, used to estimate the cosmological signal, with its uncertainty.

\section{Systematic effects}\label{sect:systs}
\subsection{Model Variance}
One of the main interests at this stage of the experiment is the effective and accurate description of systematic effects that might hinder the final measurement of the redshift drift. Since the latter relies on the production of a transmission model of the \lya forest, we ought to determine whether this procedure induces systematic uncertainties in the measurement.

In Sect.\ref{sect:model}, we built a model of SB2's \lya transmission as the average of 500 independent realisations of spline models, calibrated on physical simulations of the \lya forest at $z=3-4$, that statistically fit the data equally well. 
The variances at each pixel among the 500 independent models, $\sigma^2_{S_i}$, and among their derivatives, $\sigma^2_{S'_i}$,
can be interpreted as additional uncertainties related to the modelling procedure and used to study the presence of systematic effects from  modelling non-uniqueness in the measurement of the redshift drift.

In the Pixel-by-pixel method (see Sect.~\ref{sect:pix2pix}), the model variance term was accounted for by modifying the velocity contribution weights of each pixel (Eq.~\ref{eq:p2pweights}) as 
\begin{equation}
w_i  = \sigma_{v_{k,i}}^{-2} = \left[ \frac{c}{\lambda_i \left(\partial\bar{S}_i/\partial \lambda_i\right)}\right]^{-2}
\left[\sigma^2_{F_{k,i}} + \sigma^2_{S_i} + \frac{\left(\bar{S}(\lambda_i) - F_{k,i}\right)^2}{\left(\partial\bar{S}_i/\partial \lambda_i\right)^2}\sigma^2_{S'_i}\right]^{-1},
\end{equation}
where we propagated the model variance $\sigma^2_S$ and the model derivative variance $\sigma^2_{S'}$ to the velocity contribution variance. 

Similarly, the same variance term can be taken into account when correlating the transmission model to the data via the likelihood maximisation, adding a noise term at the denominator of Eq~\ref{eq:likelihood}, as
\begin{equation}\label{eq:likelihood_edit}
    \ln\mathcal{L}(\delta v_k) = -\frac{1}{2}\sum_{i=1}^{N_{\mathrm{pix}}} \frac{\left[F_{k,i} - \bar{S}(\lambda_i, \delta v_k)\right]^2}{\sigma^2_{F_{k,i}} + \sigma^2_{S_i}}.
\end{equation}

We performed the measurement again, following the same procedures as reported above for both methods, adopting the model non-uniqueness noise terms and recovering 
$\dot{v} = -3.18\pm3.57~\msyr$ (Pixel-by-pixel) and $\dot{v} = -3.59\pm3.72~\msyr$ (likelihood correlation) with a minimal increase of the uncertainty ($\lesssim 2\%$). The systematic uncertainty due to model variance is subdominant with respect to the statistical uncertainty and does not provide an important source of systematic uncertainty in the measurement. 
Moreover, since the transmission model is built on the combination of all exposures, in future developments of the measurement with higher S/N, the model variance is expected to decrease and remain subdominant to the photon noise, never providing a significant source of systematic uncertainty. 

\subsection{Single exposures analysis}\label{sect:Obs}

In order to investigate systematic effects and spurious velocity contribution in the single ESPRESSO exposures, we carried out the measurement described in the previous sections, comparing the \lya transmission model to the single rebinned exposures, instead of the three combined epoch spectra. As before, in both methods the model is compared to each spectrum to estimate $\delta v_k$, where $k = 1,\dots,23$, indexes the individual exposure. 
As in Sect.~\ref{sect:pix2pix}, the cosmological acceleration was estimated by performing a linear fit in the $\delta v - t$ plane across all individual exposures. The resulting estimated cosmic accelerations are $\dot{v} = -3.43\pm3.58 ~\msyr$ (Pixel-by-pixel) and $\dot{v} = -4.73\pm3.60~\msyr$ (Likelihood), both fully consistent with each other and with the measurements obtained from the combined epoch spectra.

\begin{figure}[t]
    \centering
    \includegraphics[width=\linewidth]{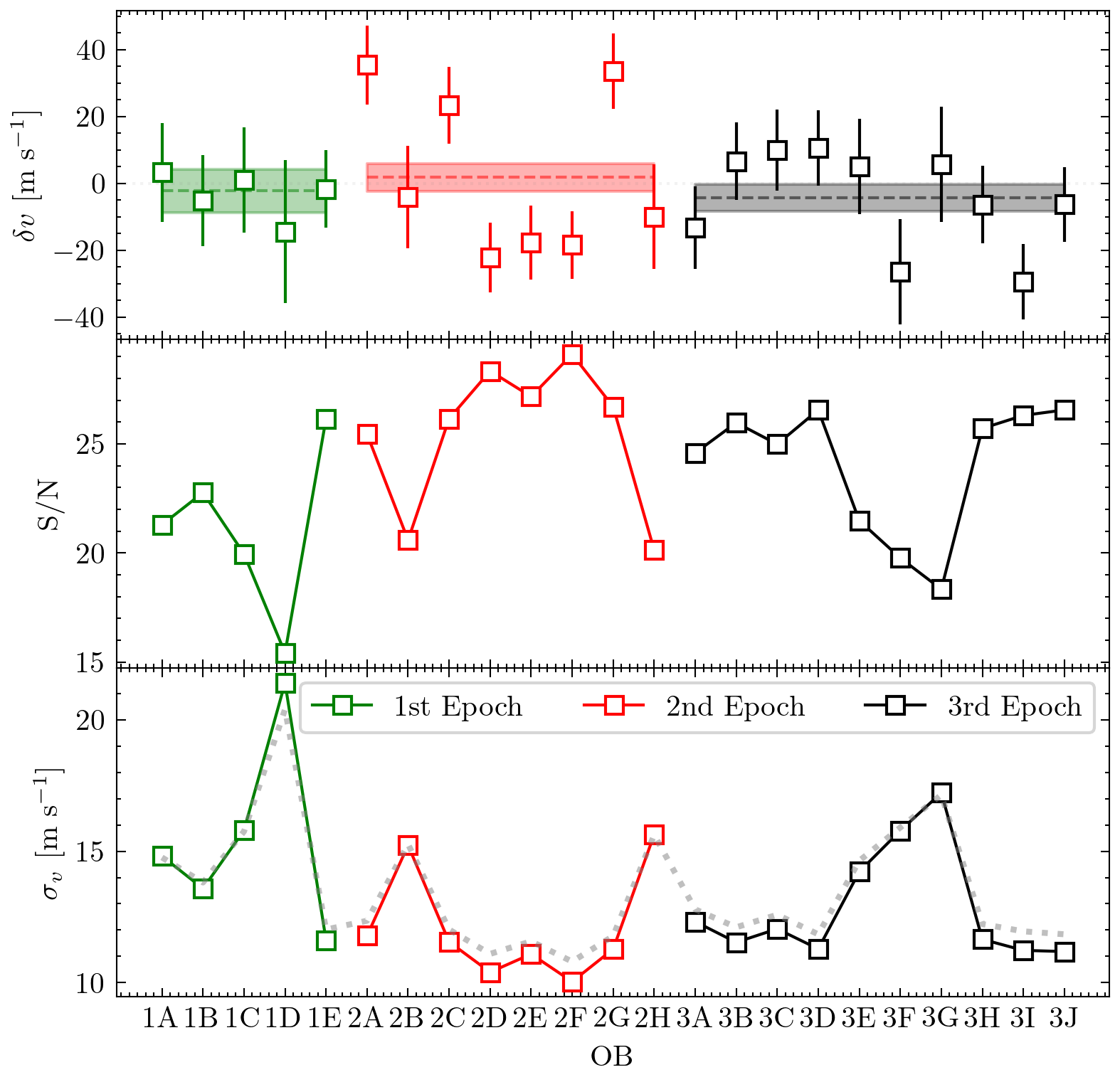}
    \caption{Top panel: Velocity shift between transmission model and single exposure flux, subdivided in the three observational epochs: first (green), second (red) and third (black). The horizontal dashed lines report the average velocity shift for each epoch, weighted by single exposure velocity shift uncertainty, and their uncertainty as shaded areas. Middle panel: median S/N per $1~\kms$ pixel at continuum in the forest for each exposure. Bottom panel: velocity shift uncertainty for each exposure. The dotted grey line reports the expected precision computed as a simple ${\rm S/N}^{-1}$ scaling.}
    \label{fig:SingleOB_snr}
\end{figure}

Fig.~\ref{fig:SingleOB_snr} shows the velocity shift between the \lya transmission model and the normalised flux of the single exposures. The shaded areas represent the weighted average velocity shift across the single epochs, with its uncertainty. The median S/N per pixel at continuum level for each exposure is reported in the figure, as well as the measurement uncertainty obtained on each spectrum.

\subsection{Wavelength calibration}\label{sect:wavelength}
Due to the small nature of the signal we are looking for ($\dot{v}\sim~0.5~\cmsyr$), a thorough assessment of systematic effects of instrumental nature is required. To this end, we evaluated the effects of the two calibration sources available on ESPRESSO on the acceleration measurement. In the previous sections of this work, we have presented the results based on the application of the FP+ThAr calibration frames, whereas in this section, we discuss their differences from LFC-based calibrations. To do so, a new complete reduction of the exposures has been carried out. We stress that both reduction processes consider a Gaussian line profile when modelling the calibration lines, instead of more complex non-parametric LSF profiles such as the one presented in \cite{Schmidt2024}, which are shown to mitigate calibration issues.

To consistently take into account variations of the wavelength solution between the two calibration frames, we re-performed the full analysis on the LFC data, including the generation of a new model. Then the model was compared to the three-epoch datasets (based on the same LFC calibration) following the same methodologies described above, yelded $\dot{v}=-2.54   \pm 3.59~ \msyr$ (Pixel-by-pixel, Sect.~\ref{sect:pix2pix}) and $\dot{v}=-2.85 \pm 3.65$ (Likelihood, Sect.~\ref{sect:Likelihood}). Both measurements are compatible with each other, and with the theoretical \lcdm acceleration expected in a \lcdm universe at $z=3.57$. Note that the acceleration measurements based on the two calibrations show a difference of $\Delta \dot{v}_{(\rm LFC - ThAr) }\sim0.9~\msyr$, i.e. the estimated drift appears slightly faster in the ThAr-calibrated spectra. This difference is subdominant to the statistical uncertainties given by the current S/N level, and we find that calibration sources do not provide dominant systematic effects at present level time. In Sect.~\ref{sect:app_wavelength_2} we investigate further the differences between FP-ThAr and LFC calibrations and the stability of their wavelength solutions over time. 

\section{Future Perspectives}\label{sect:forecast}
From the results of the three-epoch analysis, we can estimate at first order the required experiment timeline for a meaningful detection of the drift. Note however that, at this stage, the available data are limited and the extrapolated measurement uncertainty can be susceptible to non-negligible sightline-to-sightline variations, as well as yet unidentified sources of systematic uncertainty. 

\begin{figure}[t]
    \centering
    \includegraphics[width=\linewidth]{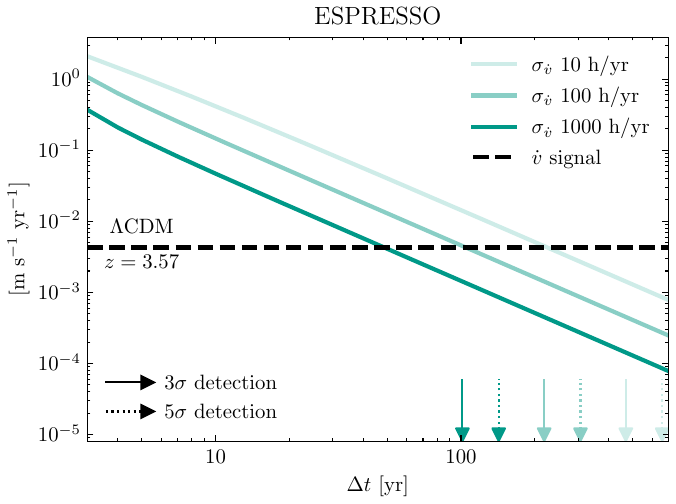}
    \caption{Extrapolation of the present results based on three epochs to a long-term ESPRESSO redshift drift experiment targeting SB2 with a continuous monitoring programme of 10, 100, and 1000 hours of integration per year. Solid green lines define the forecasted measurement precision of such a programme, depending on the allocated time, as a function of the experiment baseline. Horizontal black dashed line defines the magnitude of the cosmological signal in a \lcdm scenario at $z=3.57$, i.e. the mean redshift of SB2's forest. Vertical arrows denote the timestamp at which a 3$\sigma$ (solid) and 5$\sigma$ (dotted) detection is achieved for each time allocation, with the same colour coding.}
    \label{fig:ESPRESSO_forecast}
\end{figure}

Following the formalism of Paper I, assuming an observing strategy of $N_e$ epochs, evenly spread throughout the temporal baseline with cadence $\delta t=1~\rm yr$, so that $\Delta t = \left(N_e - 1\right)\delta t$, and with $T$ hours of integration time of SB2 per epoch, the precision of a measurement of the cosmic accelerations ($\dot{v} \sim \Delta v /\Delta t$) scales as
\begin{equation}
    \sigma_{\dot{v}} \propto g ~ {\rm S/N_{tot}}^{-1} ~ \Delta t^{-1},
\end{equation}
where ${\rm S/N_{tot}}$ is the total S/N achieved over all $N_e$ epochs and scales as ${\rm S/N_{tot}}\propto\sqrt{N_eT}$. The form factor $g$ follows from Eq.~\ref{eq:form} and, in the case of the uniform observing strategy that we assumed, takes the analytical form $g=\sqrt{3 (N_e-1)/(N_e+1)}$, with $g\approx1.7 $ for $N_e\gg1$.
Fig.~\ref{fig:ESPRESSO_forecast} shows the extrapolation of the present measurement uncertainty to a long-term experiment carried out with ESPRESSO integrating SB2 for 10, 100, 1000 hours each year, as a function of total experimental baseline. 

\begin{figure}[t]
    \centering
    \includegraphics[width=\linewidth]{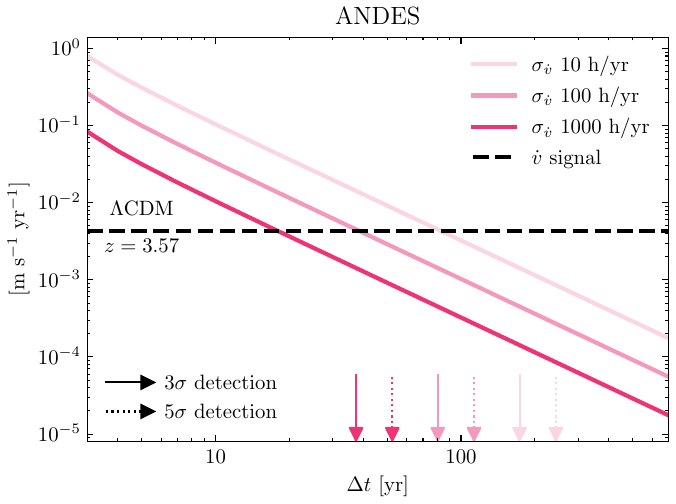}
    \caption{Extrapolation of the present results based on three epochs to a long-term ANDES redshift drift experiment targeting SB2 with a continuous monitoring programme of 10, 100, and 1000 hours of integration per year, assuming a $\epsilon_A=10\%$ efficiency.}
    \label{fig:ANDES_forecast}
\end{figure}

Similarly, we extrapolated the expected temporal baseline required for a significant detection in an ELT/ANDES-like experiment by simply considering that the ELT will have a collecting area a factor 20x larger than one VLT unit. In terms of efficiency, ANDES is expected to have similar specifications to ESPRESSO ($\epsilon_A\sim10\%$); however, this value might change during future design phases. To generalise, we can assume that one hour of ANDES integration is equivalent to $20\times\left(\epsilon_A/0.1\right)$ hours of ESPRESSO integration time, in terms of the number of photons collected by the telescope and the instrument. Thus, the temporal baselines to achieve significant detection with ANDES are shortened by a factor $20^{1/3}\times\left(\epsilon_A/0.1\right)^{1/3}\sim2.7$, as shown in Fig.~\ref{fig:ANDES_forecast}, with respect to an ESPRESSO experiment.


\begin{table}[]
    \centering
    \caption{Temporal baselines required to achieve a 1$\sigma$, 3$\sigma$, or 5$\sigma$ detection following up the SB2 spectra with a total integration time of 10, 100, or 1000 hours per year, if performed with ESPRESSO or ANDES (assuming a $10\%$ efficiency).}
    \begin{tabular}{l|rrr|rrr}
    \hline
      &  \multicolumn{3}{c|}{ESPRESSO} & \multicolumn{3}{c}{ANDES} \\
      \hline
     T      & 1$\sigma$  & 3$\sigma$ & 5$\sigma$  & 1$\sigma$  & 3$\sigma$ & 5$\sigma$ \\ 
    $[$h/yr]  & [yr]       & [yr]      & [yr]    & [yr]       & [yr]      & [yr]   \\ \hline
    10    & 225  & 469  & 659 & 83  & 174  & 243 \\
    100   & 104  & 218  & 306 & 38  & 80   & 112 \\
    1000  & 48   & 101  & 142 & 18  & 27   & 52 \\
    \hline
    \end{tabular}

    \label{tab:unce_future}
\end{table}
\subsection{A realistic observational campaign}
In a realistic scenario for a future observing campaign, ANDES will likely not begin observations before 2040. In the meantime, carrying out the programme with ESPRESSO is crucial to extend the experiment's baseline and to characterise systematic effects at the centimetre per second per year level. Moreover, a realistic scenario cannot be based solely on one target, but requires a sample of bright quasars that uniformly cover the sky, maximising observable time and reducing cosmic variance effects related to a single sightline approach. The Golden Sample (GS) of 7 bright quasars ($2.9\le z\le4.7$) presented in the QUBRICS Survey \citep{Qubrics23} was proposed specifically with this aim.

We can extend the extrapolated results presented in the previous section to a more complex, multi-target, multi-instrument campaign, assuming to continue the experiment with ESPRESSO with a yearly cadence of $T_E$ hours of integration, and follow up with $T_A$ yearly hours of ANDES integration as the instrument comes online. Moreover, assuming that the pressure on the VLTs will significantly drop after the commissioning of the ELT, we can foresee allocating a large amount of yearly time on ESPRESSO, such as 1000 h/yr across all the targets, equivalent to all the dark time of one dedicated VLT UT.
On the other hand, ELT's high demand will limit the amount of ANDES time devoted to the experiment. Here we assume an allocation of 100 h/yr (i.e. 12.5 nights per year). To account for possible changes to the final ANDES efficiency in the next stages of the instrument's design, in this section, we generalise the ANDES exposure time $T'_A=T_A\times\left(\epsilon_A/0.1\right)$, i.e. all results are given assuming $\epsilon_A=0.1$, and shall be rescaled if one wants to extrapolate the experiment to a different efficiency. 

On first approximation, we can extrapolate the results obtained on SB2 to estimate the redshift drift measurement precision obtainable with the sightlines of the other GS targets as
\begin{equation}\label{eq:sigma_gs}
    \sigma^{\rm optical}_{\dot{v}} \propto g\,{\rm S/N}^{-1} \Delta t^{-1}\left(1+z\right)^{-\gamma},
\end{equation}
where $g$ is the form factor computed following Eq.~\ref{eq:form}, $\Delta t$ is the experiment's baseline, $\gamma=1.7$ for $z<4$ and $0.9$ otherwise, and ${\rm S/N}$ is the total signal-to-noise reached by the target. Assuming shot-noise limited observations, ${\rm S/N}$ is extrapolated from the results of SB2 as
\begin{equation}\label{eq:snr_gs}
    {\rm S/N} \propto  10^{-0.2\Delta m}\sqrt{\left(\Delta t+1\right)\, T/N_{\rm obj}},
\end{equation}
where $\Delta m = m_j - m_{\rm SB2}$ is the difference in apparent magnitude between the j-th target and SB2, $\Delta t$ is the experiment baseline, $N_{\rm obj}$ is the number of targets considered. The total amount of integration time per year, $T$, depends on the time allocations of both ESPRESSO and ANDES as $T=T_E$ before 2040, and $T=T_E+20\cdot T_A'$ after 2040, assuming that we will continue ESPRESSO observations to better control systematic effects.

Note that the present analysis is based solely on the results obtained for the SB2 sightline. As shown by \citet{Liske08}, different QSO sightlines can exhibit variations in the redshift drift measurement uncertainty of up to $\sim 20\%$, due to differences in the distribution of strong and weak \lya absorption lines. 
For this reason, in Paper~II we defined the normalised velocity uncertainty
\begin{equation}
    \sigma_{100} \equiv \sigma_{v}\left(\frac{\rm S/N}{100}\right),
\end{equation}
which represents the uncertainty in the velocity positioning of a model on the target \lya transmission, assuming data with ${\rm S/N}=100$. This quantity allows us to characterize the intrinsic properties of a given spectrum and to assess each target's constraining power on the redshift drift signal.
From the results reported computed in Eqs.~\ref{eq:sdv_p2p}-\ref{eq:sdv_like}, we find that SB2 has $\sigma_{100}^{\rm SB2} \approx 2.95~\ms$, while in Paper~II we showed that SB1 (J2125) has $\sigma_{100}^{\rm SB1} \approx 3.74~\ms$. These values are used to anchor the uncertainty scaling defined in Eq.~\ref{eq:sigma_gs} for SB1 and SB2.
At present, a high-resolution analysis of the remaining five targets of the GS is still lacking, and a robust estimate of the achievable measurement uncertainties based on their \lya forests cannot yet be performed. To account for this uncertainty, we assume that the other targets have spectra providing a precision that may deviate by up to $\pm 20\%$ from $\sigma_{100}^{\rm SB2}$.

Complementary to the optical experiment, we can foresee a redshift drift experiment based on the radio observations of $z\sim0.5$ DLA \hi 21 cm absorption lines in the spectra of radio-loud background sources. To first order, we extrapolated the results of \cite{Darling2012}, based on Green Bank Telescope (GBT, \cite{GBT}) observations of 13 targets over a time span of a decade, to modern radio facilities such as the Five-hundred-meter Aperture Spherical Telescope (FAST, \cite{FAST2019}). Moreover, we can envision that a dedicated radio experiment will utilise hundreds of targets, whose discovery is a direct result of SKA prototype surveys such as ASKAP FLASH \citep{ASKAP}. The cosmic acceleration uncertainty follows from the radiometer equation and scales as
\begin{equation}\label{eq:sigma_radio}
        \sigma^{\rm radio}_{\dot{v}} \propto g \left(\frac{A_{\rm eff}}{T_{\rm sys}}\right)^{-1}N_{\rm lines}^{-1/2}~t_{\rm exp, tot}^{-1/2}~\Delta t^{-1},
\end{equation}
where $A_{\rm eff}/T_{\rm sys}$ is the sensitivity of the radio telescope, $N_{\rm lines}$ is the number of lines considered in the measurement, $\Delta t$ is the total experiment baseline, and $t_{\rm exp, tot}$ is the total integration time accumulated over $\Delta t$.

Given the expected uncertainties of the radio and optical regime, at each epoch we generated $10^4$ realisations of the experiment, randomly scattering the measurements of the 7 targets of the GS and the binned radio result around the expected \lcdm cosmic acceleration value at the corresponding redshift, adopting the estimated uncertainty as standard deviation of Gaussian noise.
Each realisation was fitted with the linear relation
\begin{equation}
    \dot{v} = \frac{d\dot{v}}{dz}(z-z_0) +  \dot{v}_{z_0},  
\end{equation}
where the pivot point $z_0=3.5$ was chosen for consistency with \citet{Qubrics23}.
From the 2D posterior distribution of the fitted values $\{\frac{d\dot{v}}{dz}, \dot{v}_{z_0} \}$ at each epoch, we measured the confidence level at which the null signal, i.e. $\dot{v}\equiv0$, is excluded. This process was repeated under the assumptions of either an ESPRESSO-only experiment with 1000 hours of integration per year, an ANDES-only experiment with 100 h/yr (and $\epsilon_A=0.1$) starting from 2040, a combined ESPRESSO+ANDES campaign with the same integration times as the latter cases, and a joint ESPRESSO+ANDES+FAST campaign where 300 \hi 21 cm lines at $z\sim0.5$ are observed yearly.

\begin{figure}
    \centering
    \includegraphics[width=\linewidth]{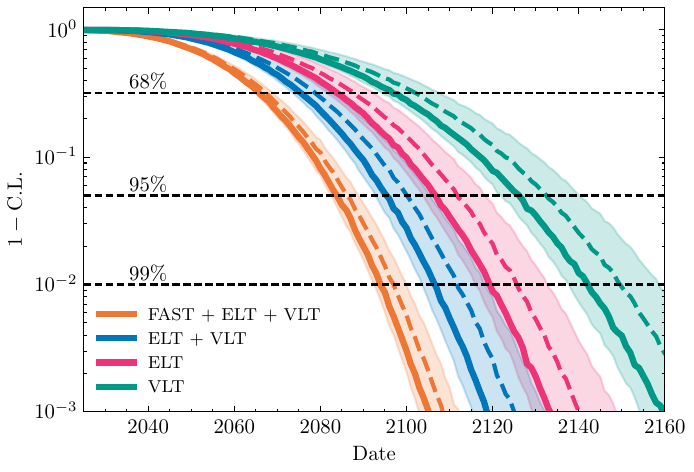}
    \caption{Dashed lines show the confidence level at which the null drift, $\dot{v}\equiv0$, is excluded, as a function of time for different experimental configurations. Green: 1000 h/yr of ESPRESSO; Magenta: 100 h/yr of ANDES starting in 2040; Blue: combination of the single ESPRESSO and ANDES cases; Orange: combination of the single ESPRESSO and ANDES cases, plus yearly FAST measurements of 300 \hi 21cm absorption lines, two hours of exposure per radio target. Shaded regions report the variation of the confidence levels under the assumption of $\pm20\%$ uncertainty on $\sigma_{\dot{v}}$ of the six targets of the GS not analysed in this work. Solid lines report the same instrumental configurations, limiting the optical measurements to only SB1 and SB2.
    Black dashed horizontal lines report the $68\%$, $95\%$, and $99\%$ confidence thresholds.}
    \label{fig:jointforecast}
\end{figure}

\begin{figure}
    \centering
    \includegraphics[width=\linewidth]{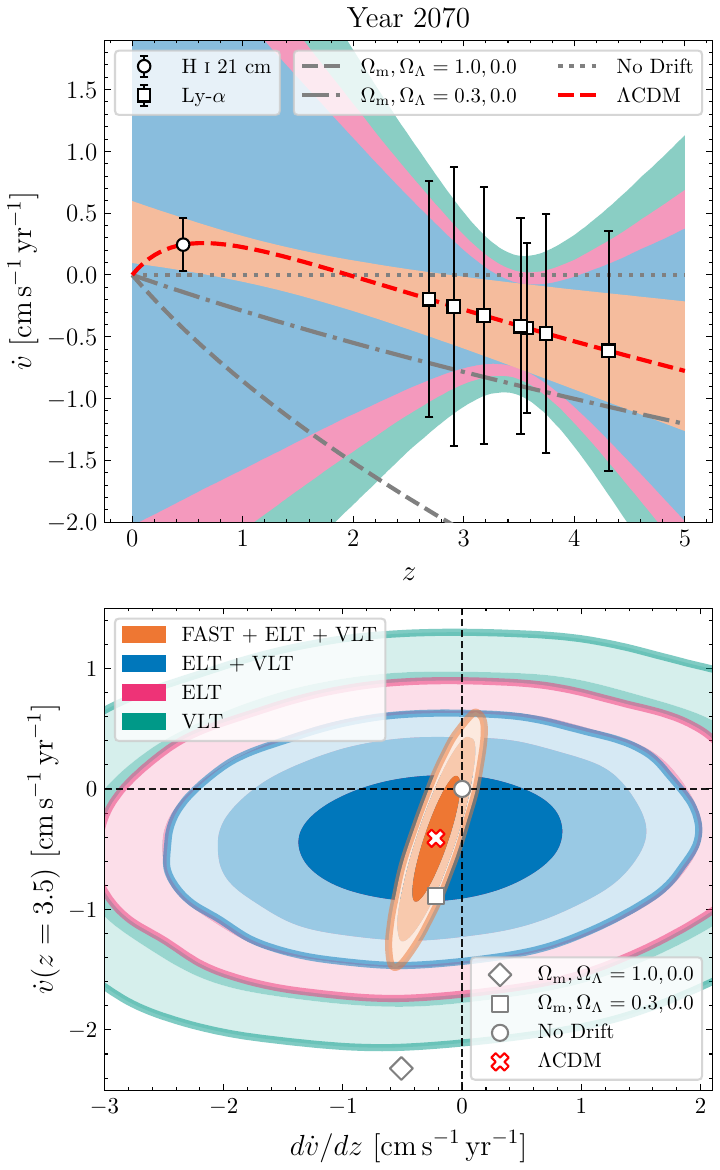}
    \caption{Extrapolated status of the experiment in the year 2070. Top panel: simulated measurements of the seven GS quasars (black squares) and the 300 H~\textsc{i} 21 cm lines (binned, black circle).  Shaded areas report the $68\%$ fit uncertainty of the $10^4$ realisations of the experiment at this epoch for the scenarios of a VLT-only experiment (1000 h/yr of ESPRESSO, green), an ELT-only experiment (100 h/yr of ANDES after 2040, magenta), an ELT+VLT campaign (combination of the latter two cases, blue), and a FAST+ELT+VLT scenario (combination of the latter case and 300 H~\textsc{i} 21 cm lines observed yearly with FAST,  orange).
    Only the acceleration uncertainties $\sigma_{\dot{v}}$ of the FAST+ELT+VLT scenario are reported as the scatter points' error bars for clarity. Red dashed line reports the expected \lcdm cosmic acceleration, whereas the grey lines describe Universes with no drift (dotted), $\Omega_{\rm m}=1.0;\Omega_\Lambda=0.0$ (dashed) and $\Omega_{\rm m}=0.3;\Omega_\Lambda=0.0$ (dash-dotted).
    Bottom panel: Contours showing the $68\%$, $95\%$ and $99\%$ levels of the fitted parameter's posterior distribution. The null-signal hypothesis ($\dot{v}\equiv0$) is defined by the white circle. White square and diamond describe the fitting parameters of the $\Omega_{\rm m}=1.0;\Omega_\Lambda=0.0$ and $\Omega_{\rm m}=0.3;\Omega_\Lambda=0.0$ Universes, respectively. \lcdm is defined with the red cross.
    In the FAST+ELT+VLT scenario, the null hypothesis is excluded at more than $68\%$ confidence level at the shown epoch.
    }
    \label{fig:YEAR2070}
\end{figure}

Figure~\ref{fig:jointforecast} shows the confidence level at which the null drift is excluded for the four different experimental configurations as a function of time, where the crossing times of the $68\%$,  $95\%$ and $99\%$ confidence levels are also reported in Table~\ref{tab:confidence}. The unknown properties of the GS sightlines are reflected in uncertainties of approximately $\sim5\%$ and $\sim8\%$ in the extrapolated detection timelines for the radio+optical (FAST+ELT+VLT) and optical-only (ELT+VLT, ELT, VLT) scenarios, respectively. Moreover, the same figure shows the effect of focusing all the optical exposure time to only the two brightest targets in the GS: J212540.96-171951.32 (SB1; $z_{\rm em}=3.897$, $i=16.548$; see Paper II) and J052915.80-435152.0 (SB2; analysed in this work). Limiting the monitoring campaign to these two bright targets, which have opposite positions in the sky and are well-suited for continuous monitoring, shortens the experimental baseline by $\sim4-7 \%$, depending on the considered scenario. 
Figure~\ref{fig:YEAR2070} shows the expected experiment status in 2070 (without cosmic variance uncertainty scaling), when a first $68\%$ non-null detection is expected to be achieved by the full FAST+ELT+VLT configuration, while targeting all 7 objects of the GS. By the same epoch, an Einstein-de Sitter Universe can be excluded at more than 99\% confidence level by all observing scenarios. 

\renewcommand{\arraystretch}{1.3}
\begin{table}[]
    \centering
    \begin{tabular}{c|lll}
    \hline
         C.L.& $68\%$ & $95\%$& $99\%$ \\
         \hline
         VLT          & $2102_{-7}^{+6}$ & $2131_{-11}^{+8}$  & $2148_{-12}^{+9}$ \\
         ELT          & $2088_{-6}^{+5}$ & $2112_{-8}^{+7}$   & $2126_{-10}^{+7}$ \\
         ELT+VLT      & $2079_{-5}^{+4}$ & $2101_{-8}^{+5} $  & $2112_{-8}^{+7} $ \\
         FAST+ELT+VLT & $2068_{-3}^{+1}$ & $2087_{-4}^{+2} $  & $2097_{-4}^{+4} $ \\
         \hline
    \end{tabular}
    \caption{Years at which the null acceleration signal is excluded at $68\%$, $95\%$, and $99\%$ confidence level, for different experimental configurations, when targeting the full QUBRICS GS. VLT: 1000 h/yr of ESPRESSO observations of the GS; ELT: 100 h/yr of ANDES starting in 2040; VLT+ELT: combination of the single VLT and ELT cases; VLT+ELT+FAST: combination of the single VLT and ELT cases, plus FAST measurements of 300 \hi 21cm absorption lines. All estimates report an uncertainty interval due to the unknown properties of the GS sightlines.}
    \label{tab:confidence}
\end{table}

\section{Result and discussion}\label{sect:results}
We have carried out the third epoch of observations of the ESPRESSO redshift drift experiment on quasar J052915.80-435152.0 (SB2), expanding on the effort reported in Paper I. With the newly added third epoch of observations, the total S/N per $1~\kms$ pixel at continuum in the \lya forest becomes $\sim 113$.

In this work, we applied the same modelling technique as presented in Paper I, based on an ensemble of 500 spline models of the \lya transmission profiles, where the spline scale is calibrated on realistic simulations of the \lya forest and tailored to the S/N of the spectra in consideration. The model obtained was used to measure the velocity drift between the three datasets with two independent methods:
\begin{itemize}
    \item Pixel-by-pixel method (Sect.~\ref{sect:pix2pix}): $\dot{v} = -3.43 \pm 3.56~\msyr$;
    \item Model Likelihood (Sect.~\ref{sect:Likelihood}): $\dot{v}=-3.63\pm3.65~\msyr$.
\end{itemize}
Both methods yield results that are compatible with each other in terms of measurement and uncertainty. The measurement precision of both methods is about 20\% larger than the theoretical prediction from the scaling relation of \cite{Liske08} (Eq.~\ref{eq:liske}; $\sigma_{\dot{v}}\approx 2.9~\msyr$). In this work, we based the drift measurement on a single sightline, whereas the scaling relation is tuned to the average result over an ensemble of QSOs and, as shown by \cite{Liske08} and in Paper II, differences in the measured velocity uncertainty are expected among sightlines, with variations as high as 20\%, due to the properties of the \lya forest (e.g. density of strong and weak absorbers, metal contaminant, DLAs) on each sightline. 
Moreover, the current data's S/N is much lower than the regime used to anchor the scaling relation, of order a few thousand, and one could expect large uncertainties on the extrapolated values from the relation at the current low-S/N regime.

As expected, with a longer baseline and a higher S/N, we pose a new constraint on the redshift drift, improving on the one presented in Paper I. The current data are not sufficient to break the best constraint found in the literature of $\sigma_{\dot{v}} = 2.2~\msyr$ coming from the analysis of 21 cm \hi lines on 10 objects over a baseline of a decade \citep{Darling2012}, although this limit is based on heterogeneous archival data at $z\sim0.09-0.69$ and is therefore complementary to our results. We estimate that 9 hours of integration of SB2 in late-2025/early-2026 (ESO period P116, run ID 116.28U4, accepted, partly executed) are sufficient to reach the same precision level. 

We extrapolated the current analysis to estimate the expected timeline of a complete experiment under different observational setups. We analysed the effect of adopting either ESPRESSO, ANDES, or both instruments to monitor the positions of \lya lines in the spectra of the seven quasars of the Golden Sample. In this framework, ESPRESSO is essential to reduce the uncertainty to a few centimetres per second per year prior to ANDES first light, enabling a robust assessment of residual systematic effects. These are likely dominated by wavelength calibration accuracy and spectral artefacts introduced during the reduction process. At the present S/N, however, such effects remain sub-dominant relative to statistical uncertainties and cannot be characterised further.
Ongoing developments in high-fidelity spectral extraction and wavelength calibration — including spectro-perfectionism \citep{Bolton2010PASP..122..248B, Piskunov2021A&A...646A..32P}, and detailed modelling of the instrumental LSF and its stability \citep[e.g.][]{Schmidt2024, Schmidt2025Iodine} — are specifically aimed at reaching the precision required for this measurement and related exoplanet applications. A dedicated treatment of these systematics is therefore deferred to future work.

At first order, a combination of both ANDES and ESPRESSO, without accounting for systematic effects nor potential future technological improvements, is expected to show first signs of significant detection by $\sim2080$. 
However, we stress the fact that these estimates are an extrapolation of the results obtained by analysing only one sightline, at relatively low S/N. \citep{Liske08} has shown that a variance of up to $20\%$ in measurement uncertainty is expected to occur between sightlines, due to the different distributions of weak and strong \lya absorbers. Since a proper high-resolution analysis of the other sightlines of the GS has not been carried out yet, the estimated timeline extrapolated from the SB2 results suffers uncertainties of about $10\%$ on the total experiment duration. Limiting the analysis to only the two brightest targets of the GS reduces the overall baseline by a factor $\sim8\%$, but might suffer from more systematics related to the reduced sample size.

Lastly, we showed that complementing the optical \lya-based experiment with a monitoring campaign of low-z \hi 21cm absorption lines performed on large modern radio facilities such as FAST greatly reduces the required experiment baseline, showing first signs of significant detection by $\sim2070$. Similarly, an extension of the current program to the bright quasars found in the northern hemisphere (e.g. J142438.12+225600.9, $z_{\rm em}=3.62$, $r=15.4~\rm mag$, \citet{Patnaik1992}) could rely on forthcoming instrumentation such as the Canary Hybrid Optical high-Resolution Ultra-stable Spectrograph (CHORUS; \citet{ZhangCHORUS2020}), under development for the 10.4m Gran Telescopio Canarias (GTC), increasing the amount of tracers of the cosmic drift and thus reducing the experiment timeline. 


\begin{acknowledgements}
The authors are thankful to J. Darling for the prolific discussion and help on radio astronomy.
AT is partly supported by the INFN INDARK and SISSA ERC IDEAS grants.
The INAF authors acknowledge financial support of the Italian Ministry of Education, University, and Research with PRIN 201278X4FL and the "Progetti Premiali" funding scheme.
This work was financed by Portuguese funds through FCT (Funda\c c\~ao para a Ci\^encia e a Tecnologia) in the framework of the project 2022.04048.PTDC (Phi in the Sky, DOI 10.54499/2022.04048.PTDC). 
CMJM also acknowledges FCT and POCH/FSE (EC) support through Investigador FCT Contract 2021.01214.CEECIND/CP1658/CT0001 (DOI 10.54499/2021.01214.CEECIND/CP1658/CT0001). CMJM is supported by an FCT fellowship, grant number 2023.03984.BD.
MTM acknowledges the support of the Australian Research Council through Future Fellowship grant FT180100194 and through the Australian Research Council Centre of Excellence in Optical Microcombs for Breakthrough Science (project number CE230100006) funded by the Australian Government.
TMS acknowledges the support from the SNF synergia grant CRSII5-193689 (BLUVES). 
The work of KB is supported by NOIRLab, which is managed by the Association of Universities for Research in Astronomy (AURA) under a cooperative agreement with the U.S. National Science Foundation.
EP acknowledges financial support from the Agencia Estatal de Investigaci\'on of the Ministerio de Ciencia e Innovaci\'on MCIN/AEI/10.13039/501100011033 and the ERDF “A way of making Europe” through project PID2021-125627OB-C32, and from the Centre of Excellence “Severo Ochoa” award to the Instituto de Astrofisica de Canarias.
ASM and JIGH acknowledge financial support from the Spanish Ministry of Science, Innovation and Universities (MICIU) projects PID2020-117493GB-I00 and PID2023-149982NB-I00.
NCS and NN acknowledge financial support by FCT - Fundação para a Ciência e a Tecnologia through national funds by these grants: UIDB/04434/2020 DOI: 10.54499/UIDB/04434/2020, UIDP/04434/2020 DOI: 10.54499/UIDP/04434/2020.
NCS is co-funded by the European Union (ERC, FIERCE, 101052347). Views and opinions expressed are however those of the author(s) only and do not necessarily reflect those of the European Union or the European Research Council. Neither the European Union nor the granting authority can be held responsible for them. This work was supported by FCT - Fundação para a Ciência e a Tecnologia through national funds by grants reference UID/04434/2025.
FP would like to acknowledge the Swiss National Science Foundation (SNSF) for supporting research with ESPRESSO through the SNSF grants nr. 140649, 152721, 166227, 184618 and 215190. The ESPRESSO Instrument Project was partially funded through SNSF’s FLARE Programme for large infrastructures.
\end{acknowledgements}

\section*{Affiliations}
\noindent
$^1$SISSA --  International School for Advanced Studies, Via Bonomea 265, I-34136 Trieste, Italy\\
$^2$INAF -- Osservatorio Astronomico di Trieste, Via G.B. Tiepolo, 11, I-34143 Trieste, Italy\\
$^3$Dipartimento di Fisica dell’Università di Trieste, Sezione di Astronomia, Via G.B. Tiepolo, 11, I-34143 Trieste, Italy\\
$^4$INFN -- National Institute for Nuclear Physics, via Valerio 2, I-34127 Trieste\\
$^5$Centro de Astrofísica da Universidade do Porto, Rua das Estrelas, 4150-762 Porto, Portugal\\
$^6$Instituto de Astrof\'isica e Ci\^encias do Espa\c{c}o, Universidade do Porto, CAUP, Rua das Estrelas, 4150-762 Porto, Portugal\\
$^7$Departamento de F\'isica e Astronomia, Faculdade de Ci\^encias, Universidade do Porto, Rua do Campo Alegre, 4169-007 Porto, Portugal\\
$^8$IFPU -- Institute for Fundamental Physics of the Universe, via Beirut 2, I-34151 Trieste, Italy
$^9$Hamburger Sternwarte, Universität Hamburg, Gojenbergsweg 112, D-21029 Hamburg, Germany11
$^{10}$European Southern Observatory (ESO), Karl-Schwarzschild-Str. 2, 85748 Garching bei Munchen, Germany\\
$^{11}$INAF -- Arcetri Astrophysical Observatory, Largo E. Fermi 5, I-50125 Florence, Italy\\
$^{12}$Instituto de Astrof\'{\i}sica de Canarias (IAC), Calle V\'{\i}a L\'actea s/n, E-38205 La Laguna, Tenerife, Spain\\
$^{13}$Departamento de Astrof\'{\i}sica, Universidad de La Laguna (ULL), E-38206 La Laguna, Tenerife, Spain\\
$^{14}$Centre for Astrophysics and Supercomputing, Swinburne University of Technology, Hawthorn, Victoria 3122, Australia\\
$^{15}$Instituto de Astrof\'isica e Ci\^encias do Espa\c{c}o, Faculdade de Ci\^encias da Universidade de Lisboa, Campo Grande, PT1749-016 Lisboa, Portugal\\
$^{16}$Departamento de Física da Faculdade de Ciências da Universidade de Lisboa, Edifício C8, 1749-016 Lisboa, Portugal\\
$^{17}$Observatoire Astronomique de l'Universit\'e de Gen\`eve, Chemin Pegasi 51, CH-1290 Versoix, Switzerland\\
$^{18}$Center for Space and Habitability, University of Bern, Gesellschaftsstrasse 6, Switzerland\\
$^{19}$Cerro Tololo Inter-American Observatory/NSF NOIRLab, Casilla 603, La Serena, Chile\\
$^{20}$INAF -- Osservatorio Astronomico di Padova, Vicolo dell'Osservatorio 5, I-35122, Padova, Italy\\
$^{21}$INAF -- Osservatorio Astrofisico di Torino, Via Osservatorio 20, I-10025 Pino Torinese, Italy\\
$^{22}$Centro de Astrobiología, CSIC-INTA, Camino Bajo del Castillo s/n, 28602 Villanueva de la Cañada, Madrid, Spain\\
\bibliographystyle{aa}
\bibliography{AT}
\appendix
\section{Wavelength Calibration Stability}\label{sect:app_wavelength_2}
Accurate and stable calibration sources play a crucial role in the measurement of tiny effects, such as the cosmic redshift drift and are expected to be the main origin of systematic uncertainty in such an experiment. 
In this section, we look for differences in the wavelength solutions provided by the FP+ThAr and LFC calibration frames. 

\begin{figure*}
    \centering
    \includegraphics[width=\linewidth]{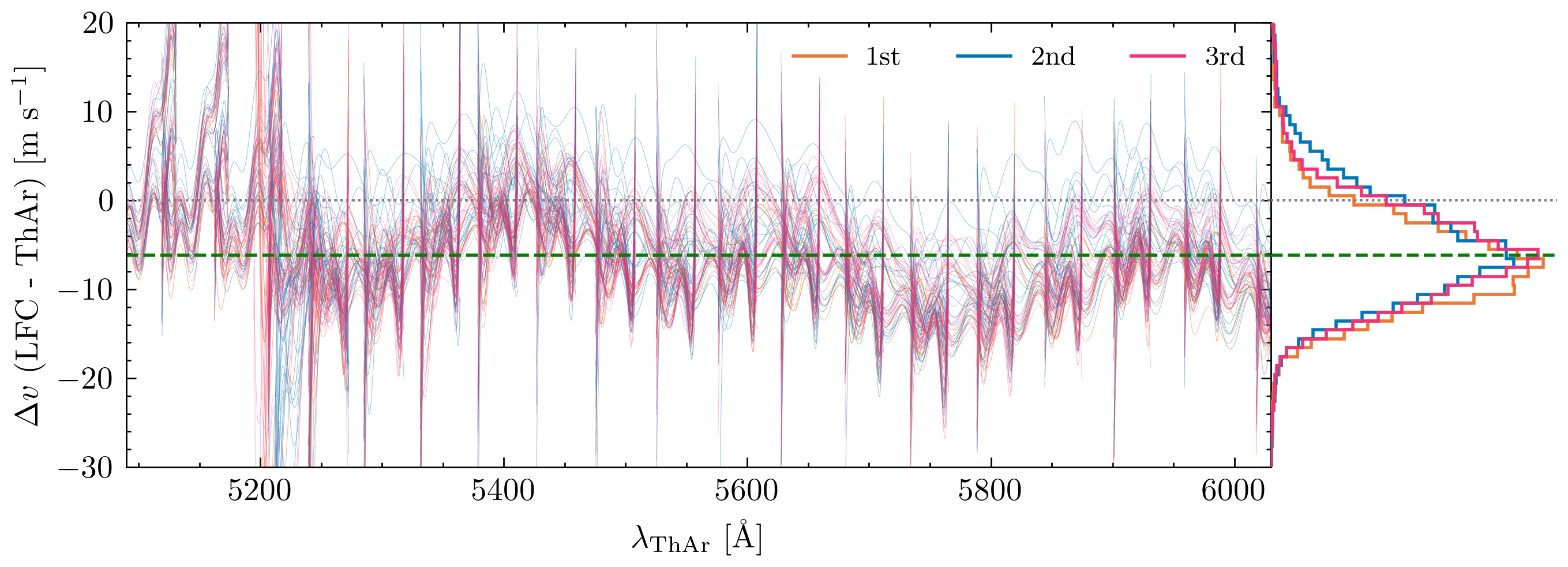}
    \caption{Velocity difference per pixel in the wavelength solutions of the order-by-order ESPRESSO spectra with LFC and ThAr+FP calibrations, as a function of ThAr-calibrated wavelength. All SB2 exposures are stacked and colour-coded by epoch: first (orange), second (blue), and third (pink). Right-hand histogram shows the distribution of the pixel velocity differences, with the same colour code. Dashed green line reports the overall median velocity difference.}
    \label{fig:calib}
\end{figure*}

We compared the wavelength solutions of each pixel of the order-by-order ESPRESSO spectra of each quasar exposure provided by the two calibration sources, as obtained through the reduction pipeline where we assumed a Gaussian LSF for both sources.
Fig.~\ref{fig:calib} shows the difference (in velocity units) between the LFC and ThAr solutions as a function of ThAr-based wavelength within the \lya forest, where large-scale oscillations are found, as well as single-order effects, especially at order edges. A large fluctuation is found in correspondence with the dichroic split between the two arms of the spectrograph (at $\sim 5200~\rm \AA$).
The median velocity difference between the two calibration sources is $-6~\ms$, consistently among the three epochs of observation.

Since the measurement of the redshift drift is intrinsically a relative problem, i.e. it measures the difference in line positions over time, a constant systematic shift between the wavelength solutions of the two calibration sources should not affect the results, as long as the analysis is carried out self-consistently on only one source.
We therefore analysed the time evolution of the median differences between LFC and ThAr solutions, $\overline{\Delta v}$, and found a trend with slope $d\overline{\Delta v}/dt = 0.36\pm0.49~\msyr$. This result, although compatible with a null time evolution of the differences in wavelength calibration,  accounts for about half of the discrepancy found between the ThAr and LFC-based drift measurements ($\Delta\dot{v}_{(\rm LFC-ThAr)}\sim0.9~\msyr$; see Sect.~\ref{sect:wavelength}).

\begin{figure}
    \centering
    \includegraphics[width=\linewidth]{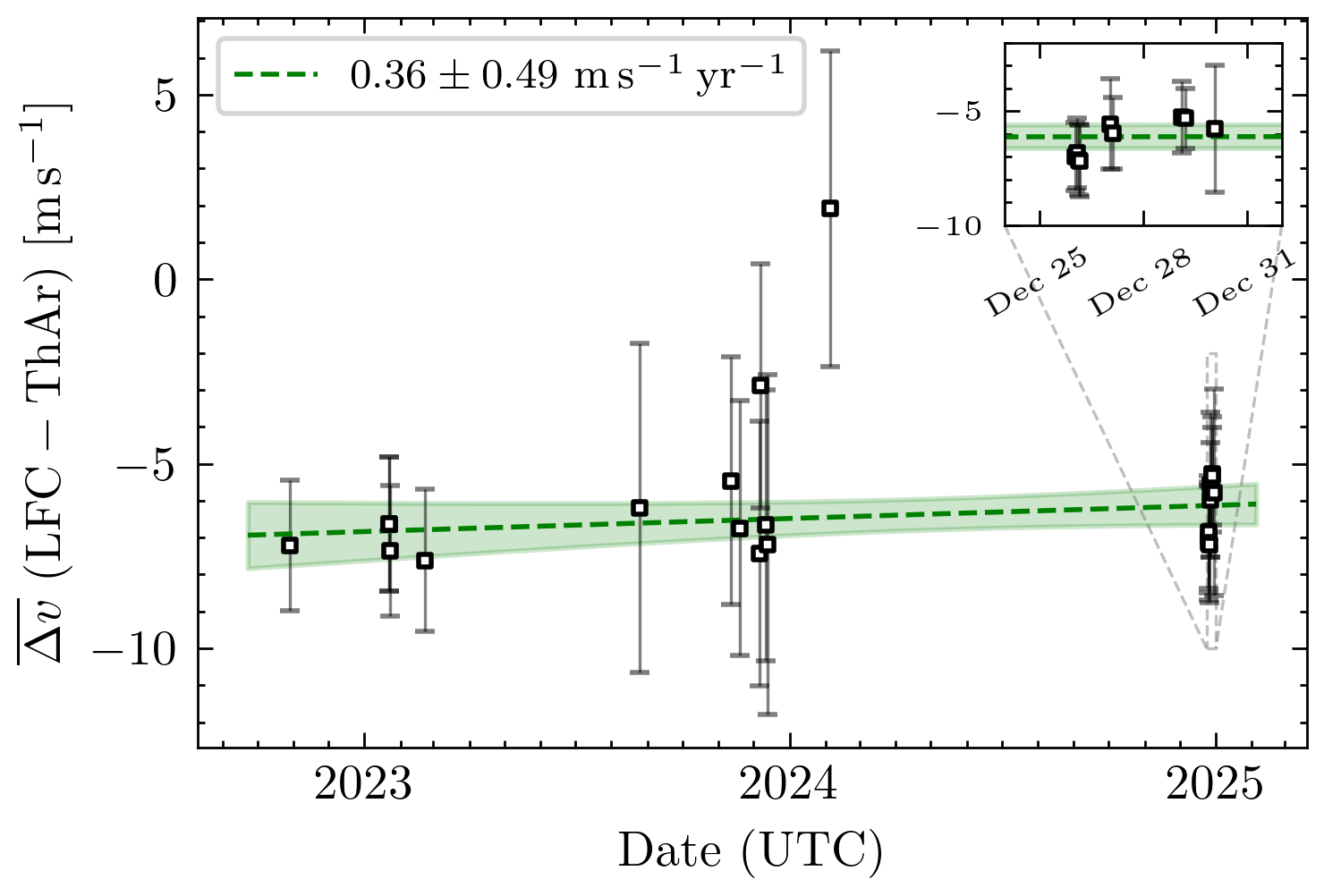}
    \caption{Time evolution of the median velocity difference in the wavelength solution of LFC and ThAr calibrated data. Green dashed line and shaded area represent the best-fit linear relation with its $1\sigma$ uncertainty region.}
    \label{fig:calib_stability}
\end{figure}
\end{document}